# Preliminary estimation of the footprint and survivability of the Chelyabinsk Meteor fragments

Cristina Parigini[1], Juan Luis Cano[2], Rodrigo Haya-Ramos[3]
*Elecnor Deimos, Tres Cantos, Madrid, 28760, Spain*

**There are several differences between the planetary entry of space vehicles and that of asteroids. In this work we do investigate the applicability of classical methods and approaches developed for debris analysis to asteroid entry. In particular, the in-house DEBRIS tool, which has been designed and developed to address the debris problem for uncontrolled re-entry objects, is used here to predict the survivability and the ground footprint of asteroid fragments. The results obtained for the Chelyabinsk event are presented as test case. A comparison with the current available information is provided, proving the validity of the proposed approach.**

## I. Introduction

The atmospheric flight of objects entering at hypervelocity is characterized by a set of complex and coupled physical phenomena such as hypersonic aerodynamics, heating, ablation, fragmentation, fragments interaction, and airburst. These objects are characterised by very high kinetic energy levels, so that aerodynamic forces and heat fluxes can produce a massive disruption combined with a high level of fragments ablation.

Sophisticated physical models describing the motion and the disruption of a meteoroid exist, and two main approaches can be distinguished. The first one is based on a hydrodynamical approximation: in this case the object is modelled as strength-less liquid-like object or drop. This approximation is applicable in case of initially continuous and totally melted impactors, which holds for very small objects. It can also be applied to the study of the motion of a dense debris cloud or compactly packed sand objects. This condition is usually related to totally disrupt large bodies, where the swarm of fragments moves together and can be represented by a single-body. Several studies have been made in this area, e.g. by Shuvalov et al.[1] and by Artemieva and Pierazzo[2]. The second approach is based instead on the analysis of progressive fragmentation of a finite number of interacting fragments, as shown by Artemieva and Shuvalov[3] and by Bland and Artemieva[4].

Simpler and faster models exist too. In this frame the pancake model, initially proposed by Chyba[5], is the most common approach. The basic idea is that the fragmented impactor expands laterally under the differential pressure between the front and back surfaces, resulting in a shape similar to a pancake. The Earth Impact Effects Program (EIEP)[6] is for example based on this approach.

In this work we use the DEBRIS tool, which has been developed and used by Elecnor Deimos in the frame of several ESA projects. Safety is an integral part of the atmospheric Mission Analysis and Flight Mechanics portfolio for the validation of mission scenarios. For this reason Deimos has a significant expertise in support to system and safety teams for the assessment of the break-up, fragmentation, re-contact analysis and debris footprint for launcher stages (ex: Vega [7]), vehicle in failure modes (ex: IXV, AREV [8], HSTS), planetary probes carriers (ex: Exomars), Resource and Propulsion Modules (ex: ARV[9, 10], HSTS), etc.

---

[1] Project Engineer, Flight Mechanics Division, Flight Systems Business Unit, cristina.parigini@deimos-space.com
[2] Head of Mission Analysis and Navigation Division, Flight Systems Business Unit, juan-luis.cano@deimos-space.com
[3] Head of Flight Mechanics Division, Flight Systems Business Unit, rodrigo.haya@deimos-space.com

DEBRIS is the acronym for the tool within the in-house Planetary Entry Toolbox [11, 12] that estimates the footprint on ground of the debris of an uncontrolled re-entry object. The purpose of this tool is to give a first shot of the impact area of the debris produced by a vehicle break-up during its atmospheric entry, exploring also the survivability of the elements. It is based on engineering models which allows for fast analyses, as risk assessment during mission design, or a statistical analysis of the main events (final state, break-up point …). DEBRIS has been designed and developed for the entry of spacecraft. However, in the frame of ESA's NEO Impact Effects and Mitigation Measures (SSA-SN-VII) study, it has been investigated how this tool can be improved and generalized for fast analyses of the entry of asteroids.

There are several differences between the planetary entry of space vehicles and that of asteroids, which make classical methods and approaches developed to address the debris problem not always directly applicable to the entry of an asteroid. First of all the knowledge of the object properties and entry conditions is limited and characterised by large uncertainty. Furthermore, the correlations used to predict the thermal loads that are usually valid for the entry of a space vehicle are not applicable for the typical range of velocities of asteroids (up to 70 km/s). Finally, the mechanism of the fragmentation is quite different, being related more to mechanical loads than thermal ones in case of asteroids. Nevertheless, thanks to its flexibility in the inputs and models definition and the possibility of managing uncertainties and worst cases, the DEBRIS tool is applied here to describe the entry and fragmentation of asteroids. Actually, only minor modifications have been implemented to run the presented simulations: they are mainly related to the inclusion of a new mass loss ablation model suitable for the entry of asteroids and to the post-processing of the results.

Based on DEBRIS tool, an estimation of the debris footprint of the Chelyabinsk meteor and a preliminary analysis of the survivability of its fragments is reported, providing also a comparison with the available information about the recovered fragments.

## II. DEBRIS tool capability

The DEBRIS tool is based on an object-oriented approach. This means that the break-up of an entry vehicle is assumed at a specific point of its trajectory implying the total collapse of the structure due to thermo-mechanical loads. After this event, each single fragment is analysed independently. The core of the tool is the simulation of the entry trajectory. Simulations are based on the Endo-Atmospheric Simulator (endoSim) within the Planetary Entry Toolbox, in which all of the vehicle and environmental models, as well as simulation options, are user-defined. To deal with uncertainties, parametric search or Monte Carlo approaches are employed.

A brief description of the tool capability follows. Some consideration on the applicability and the modifications introduced to model the entry of an asteroid are also included.

Considering debris assessments 3-DOF simulations (position and velocity) are usually suitable to represent the **re-entry dynamics** of the vehicle down to the breakup point and those of the fragments down to the demise altitude, or ground. The fragments are likely to be tumbling bodies and therefore they are modelled as ballistic low-lift objects.

The **aerothermodynamics** of a vehicle is another key point in the trajectory computations and debris assessments. It determines the drag profile, which drives the thermo-mechanical loads acting on the entry vehicle and therefore its breakup. Concerning the fragments, the final kinetic energy and possible demise altitude are strictly related to their deceleration profiles. Therefore, basic profiles of the drag coefficients depending on the regime can be assumed or a full aerodynamic characterization depending also on the vehicle configuration, the attitude, the angular rates, and possible active surface deflections.

Thermal flux estimations are usually based on empirical or semi-empirical laws, as those of Tauber for the convective heat flux [13]. However, in case of high-speed entry, both convective and radiative heat fluxes are modelled and possible coupling effects can be also considered. Based on its range of validity, a suitable model for each problem has to be identified by the analyst. It is important to notice that such models are not applicable in case of the entry of asteroids mainly because of the high entry velocities (above 14 km/s). For this reason, the thermal flux estimations and the thermal model usually applied to the entry of spacecraft produce inaccurate predictions. A simple, but more suitable, thermal model has been therefore implemented and included in the simulations.

The **breakup** represents the total collapse of the object and it is usually based on thermo-mechanical loads. In particular, pertaining to the fragmentation of the asteroids, the most common assumption is to base the breakup on a mechanical criterion.

After the breakup, the distribution of mass and dimension of the fragments can be based on a detailed debris catalogue or on statistical distributions, and trajectories are run down to ground without modelling further fragmentations. The fragments are then filtered based on thermo-mechanical loads and energy criteria to identify those that reach ground. Provided that the fragment properties generated by the disruption of an asteroid are not known, random samples are generated to explore the effects in terms of ground or demise condition of each fragment independently from the others.

## III. Simulation cases setup

On February 15$^{th}$ 2013 a small asteroid entered into the Earth atmosphere over the Chelyabinsk region in Russia: this object was not detected until its entry in the atmosphere. Even if characterized by large uncertainty, the first reconstructions agree on a 15-20 meters object with a mass of approximately 11000 tons, entering at velocity around 18 km/s. Fragmentation occurred at 30-70 km and airburst at 15-25 km [14]. The trajectory path and the fragmentation and airburst altitudes have been estimated mainly based on the analysis of several videos recording the trail left by the meteor [15,16]. Additional information has been extracted by seismic and infrasound registrations of worldwide networks of sensors and from the recovered fragments. Numerous fragments have been recovered near the Chebarkul Lake, where a hole of approximately 6 m has been found in its frozen surface (the connection with the event is still to be verified), and all along its path. The analysis of the fragments confirmed the nature of the object providing also information about its composition [17].

Based on the currently available information about the Chelyabinsk meteor event, three simulations have been run. The models and settings considered are described in the following, and a summary of the numerical values considered is provided in Table 1 and Table 2.

Concerning the re-entry dynamics, 3-DOF simulations have been run for both the entry object and the fragments. Initial conditions, reported in Table 1 in terms of position and velocity, come from the first reconstructions provided in [18] (*Cases 1* and *2*) and in [16] (*Case 3*). In particular, for *Case 3* the initial flight path angle and azimuth have been estimated to result in the given values at the end of the trajectory. The three cases cover a large range of entry velocity, between 13.4 km/s and 19.6 km/s, bracketing the values reported in other references [14,19,17]. The meteoroid trajectory has been always described as a shallow westwards path.

Table 1   Main simulation setup parameters

| Parameter | Value | | |
|---|---|---|---|
| Case | 1 [18] | 2 [18] | 3 (from [16]) |
| Altitude (km) | 32.47 | 46.75 | 91.83 |
| Longitude (deg) | 62.06 | 62.35 | 64.27 |
| Latitude (deg) | 54.92 | 54.81 | 54.51 |
| Velocity (km/s) | 13.43 | 19.65 | 17.50 |
| Flight Path angle (deg) | -16.33 | -19.73 | -18.39 |
| Heading angle (deg) | 271.60 | 276.48 | 282.41 |

The initial mass and size of the object are unknown but estimations range from 15 m to 20 m for an approximately 11000 ton mass [19,17]. Recovered fragments have been found to be composed by ordinary chondrites [17]. Therefore, an 18 m diameter spherical object of stone material has been assumed and typical values for the material properties are used. In particular, a low material strength is considered for the initial entry object representing a fragile or already fragmented internal structure.

For what concerns the fragments, a wide range of variability in terms of mass, size and density is assumed. Furthermore, a higher threshold on material strength (with respect to the entry object) has been defined to identify possible further fragmentations. A slightly lifting capability (max lifting to drag ratio of 0.1) is also considered.

For each initial velocity case, a single entry trajectory is simulated down to the breakup condition, followed by 2000 dispersed trajectories of the fragments. Uniform distributions are assumed to define the properties and size of the fragments with the aim to explore all the potential fragments that could be generated. Fragments are finally post-processed and filtered based on stagnation pressure and ablation criteria.

**Table 2    Main simulation setup parameters**

| Parameter | Value |
| --- | --- |
| Entry Object Properties | |
| Initial Diameter | 18 m |
| Initial Mass | 11000 ton |
| Material Density | 3.6 g/cm$^3$ |
| Material Strength | 10 MPa |
| Heat of Ablation | 8 10$^6$ J/kg |
| Drag Coefficient | 0.66 |
| Fragments | |
| Initial Diameter | [5 cm; 10 m] |
| Initial Mass | [0.2 kg; 2000 ton] |
| Material Density | 3.6±20% g/cm3 |
| Material Strength | 15 MPa |
| Drag Coefficient | [0.5; 1.5] |
| Lift to Drag Ratio | [0; 0.1] |
| Environment | |
| Atmosphere | USSA1976 |
| Gravity | Point Mass |

## IV. Discussion of the results

The profiles of stagnation pressure and kinetic energy as function of the altitude are reported in Figure 1. The breakup occurs between 22.8 km and 27.9 km and, as expected, higher velocities imply an earlier fragmentation (that is at higher altitude). These values are in line with the reconstruction reported in [18], but are slightly lower if compared to [16], where the fragmentation starts at 32 km when 4 MPa of dynamic pressure is reached. In other references [14, 17], fragmentation point is indicated at even higher altitudes, between 30 km and 70 km, while an altitude 23.3 km is given as the peak brightening or airburst point. In any case, given that the fragmentation point is assumed here as a single event along the trajectory it is reasonable to relate this point to the moment where the fragments start to be independent better than to the point where the fragmentation starts. In addition, the estimated kinetic energy is of the same order of magnitude of that reported in [14,17] (440 kton TNT).

The DEBRIS prediction of the groundtrack and of the fragment footprint is provided in Figure 2. The results are plotted over a Google Maps image of the Chelyabinsk region of Russia. The trajectory estimation results in a fragmentation around the Korkino city. Extending the trajectories down to ground and assuming no fragmentation lead to a landing point very close to the Chebarkul Lake for *Cases 1* and *2*: this result is consistent with the fact that the initial conditions are taken from the reconstruction made by Zuluaga and Ferrin [18], in which the Chebarkul Lake is assumed as the impact point. Concerning *Case 3*, the trajectory flies over the Chebarkul Lake reaching a region westwards of the city of Miass. This seems more in line with the refined predictions provided by Zuluaga et al. in [20] (where a review of the first reconstruction attempt [18] is made) and with the considerations reported in [21]. However, due to few degrees of difference in the initial heading angle, the final point predicted in [20], plotted in red, is approximately 10 km northern than that obtained in *Case 3*. The peak brightness point reported in [17], plotted in blue, is not far (but southern) from the predicted breakup points. The footprint on ground of the debris covers a wide area that follows the same path and reaches distances up to 15 km normal to the path of the not-fragmented object and up to 100 km far from the breakup point. Potential fragments are predicted to land approximately between the cities of Korkino and Miass.

The mass, dimension, velocity, and energy of the fragments at landing are shown in Figure 3 as function of the downrange from the breakup point. The simulations result in fragments ranging different orders of magnitude in both mass and dimension. The smallest fragments, below 1 cm, can fall along the whole path from right after the breakup up to almost 90 km away. Instead, larger fragments are likely to fall at longer downrange. The largest fragment estimated in correspondence of the Chebarkul Lake, is of almost 50 cm for *Case 2* and *Case 3* and around 2 m for *Case 1*; the mass is approximately 200 kg for *Case 2* and *Case 3* and about 10 ton for *Case 1*. This large variability in the values is strictly related to the very different initial velocity: *Case 2* and *Case 3* are characterized by higher velocity, above 18 km/s, which implies higher mass losses due to ablation, and consequently a significant reduction in the fragment size. The velocity in these cases is closer than that of *Case 1* to the most agreed range (16-18 km/s). Almost all the fragments reach the terminal velocity that varies between 30 m/s and 300 m/s, for a final energy ranging different order of magnitudes. It is important to remind that in the presented analysis the simulated fragments have been filtered based on stagnation pressure and ablated mass. This means that fragments that are likely to be further fragmented are excluded: it is expected that they will populate the low mass, high velocity, and high downrange region of the explored domain. Thus, simulation of further fragmentations can play an important role provided that the identification of high energy fragments reaching ground could be missed here.

The exact position and mass of the recovered fragments are not available; however in [14] it is stated that several fragments from 1 to 5 cm have been found in the Chebarkul Lake area; other small pieces have been found in the area near the village of Deputatskoye. A 10 cm fragment around 1 kg has also been found [22]. The blue points in Figure 2 show instead the location of three possible recovered fragments taken from [23]. All these information are completely consistent with our results in terms of fragment footprint on ground, mass, and dimension. Concerning the hole in the frozen surface of the Chebarkul Lake, from the reconstruction reported in [16], the largest mass fragment between 200 kg and 500 kg (corresponding to an object of approximately 50-60 cm) is predicted to land in this area. The same values are also reported in [24]. Once again, the DEBRIS predictions are in line with these estimations.

As an order of magnitude, each run requires less than 30 minutes (on a commercial laptop).

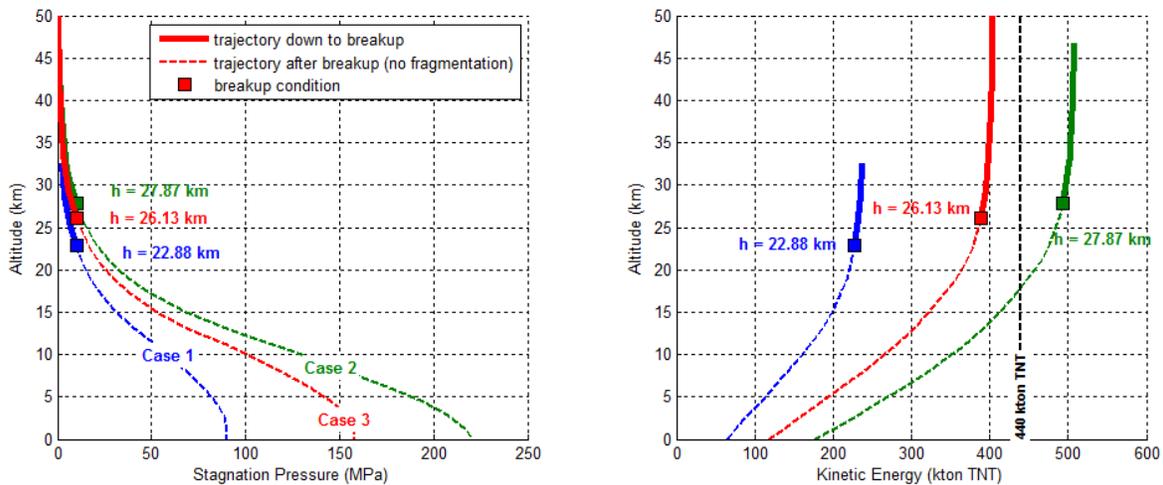

**Figure 1 Stagnation pressure and kinetic energy profiles as function of the altitude**

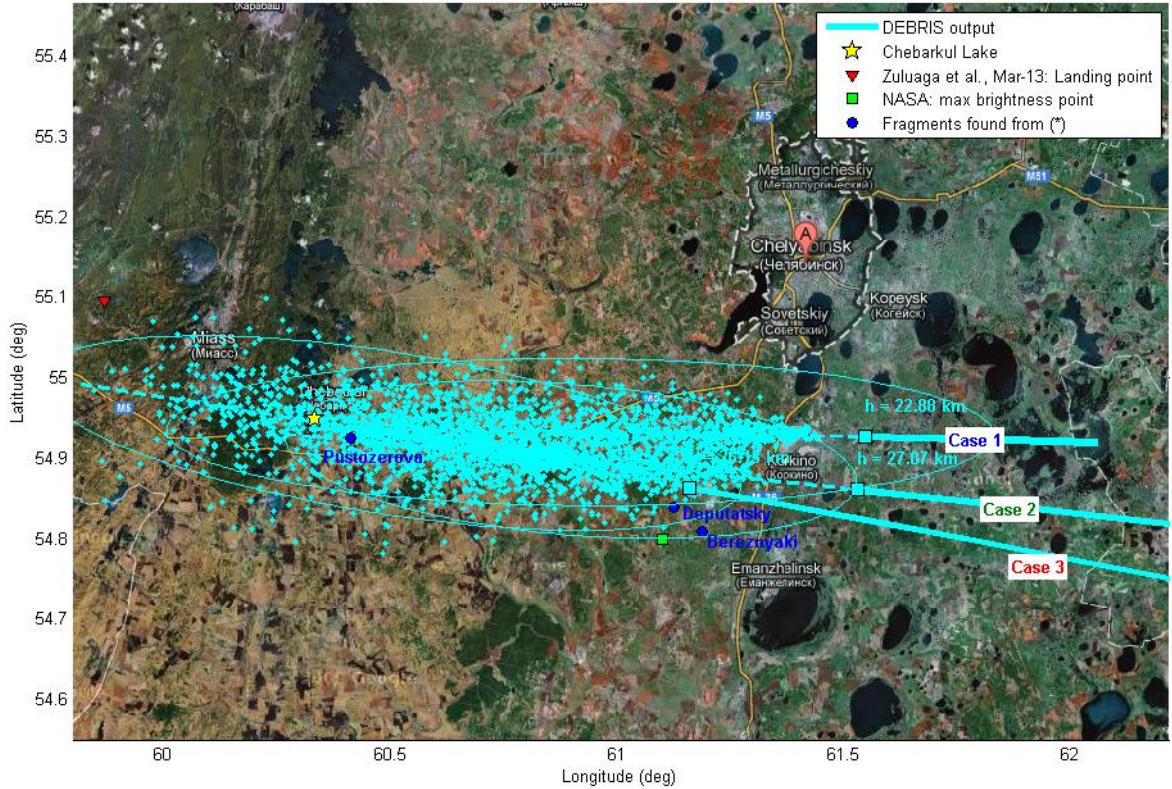

**Figure 2 Groundtrack and fragment footprint over a Google Maps image of the Chelyabinsk region**

(*) Recovered fragments as reported in [23]

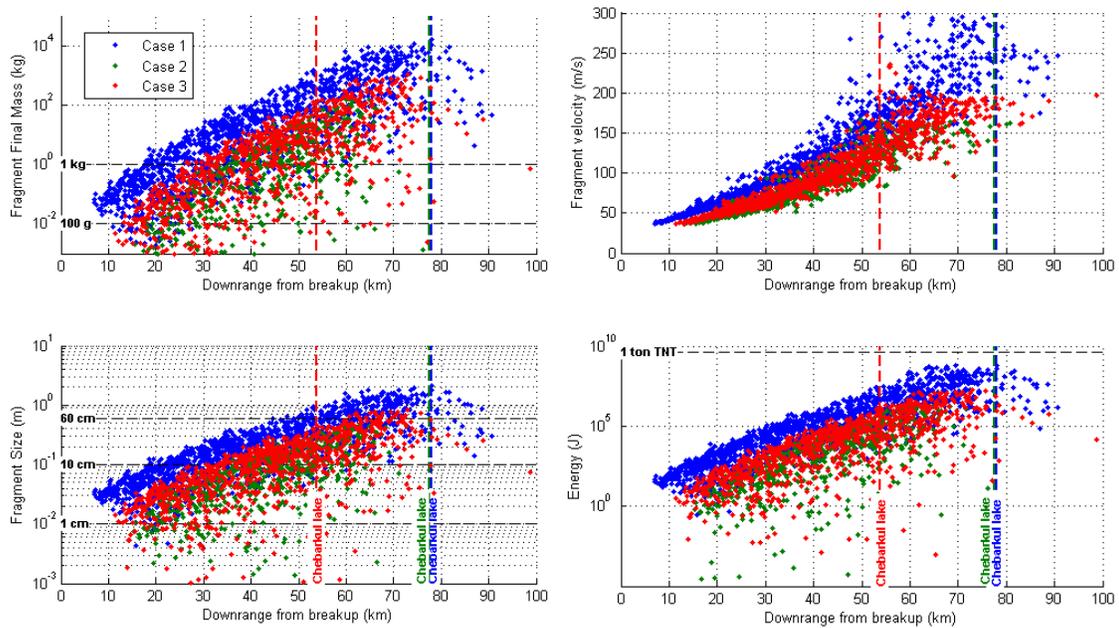

**Figure 3 Fragment final properties as function of the downrange from the breakup point; Chebarkul Lake is also represented for comparison purposes**

## V. Conclusion

The use of the DEBRIS tool applied to the atmospheric entry of an asteroid is discussed. This tool has been developed to provide fast estimations of the survivability and the footprint on ground of the debris of an uncontrolled re-entry object. The key aspects of the entry and fragmentation of a space vehicle in the Earth atmosphere, as managed by DEBRIS, have been analysed focusing on the significant differences with respect to asteroids. As a result, the tool is able to deal with both problems by setting properly the simulations inputs and by introducing only minor modifications to the code. In particular, a new mass loss ablation model suitable for the entry of asteroids has been included.

As a test case, the analysis of the Chelyabinsk event in terms of breakup and fragment footprint and survivability is reported, demonstrating the validity of the approach proposed. The predictions agree with several different reconstructions of the same event and with the available information on the recovered fragments.

Some limitations of the results presented and the areas in which the tool should be improved and generalized are also identified. They involve the coupling between the dynamics and the thermal analyses, the ulterior fragmentation after the main breakup, the inclusion of probability models for the generation of the fragments and the modelling of the airburst.

## Acknowledgments

This study has been carried out in the frame of the study entitled "NEO Impact Effects and Mitigation Measures" funded by the European Space Agency (ESA) Space Situational Awareness Preparatory Programme (SSA) and within its NEO Segment, contract number 4000106175/12/D/MRP.

## References

[1] V. V. Shuvalov, N. A. Artemieva, I. B. Kosarev, "3D Hydrodynamic Code SOVA for multimaterial flows, application to the Shoemaker-Levy 9 Comet impact problem", International Journal of Impact Engineering, 23 (1), Part 2, 847-858, 1999

[2] N. Artemieva, and E. Pierazzo, "The Canyon Diablo impact event: Projectile motion through the atmosphere", Meteoritics & Planetary Science 44, Nr 1, 25-42, 2007

[3] N. A. Artemieva, V. V. Shuvalnov, "Motion of a fragmented meteoroid through the planetary atmosphere", Journal of Geophysical Research: Planets (1991–2012) Volume 106, Issue E2, pages 3297-3309, 2001

[4] Bland, P. A., and N. A. Artemieva, "The rate of small impacts on Earth", Meteoritics & Planetary Science 41, 607-631, 2006

[5] Chyba, C.F., et al., "The 1908 Tunguska explosion: atmospheric disruption of a stony asteroid", Nature, 361, p.40.

[6] Collins, G. S., Melosh, H. J., Ivanov, B. A., "Modeling damage and deformation in impact simulations", Meteoritics and Planetary Science, 39, 217-231, 2006

[7] D. Bonetti et al. BLAST Mission Analysis, Flight Mechanics and GNC. 3rd International ARA Days, Arcachon, France, May 2011

[8] R. Haya et al. Assessment Of Vehicle Concepts For Space Transportation And Re-Entry Experimental Missions. 1st International ARA Days, Atmospheric Reentry Systems, Missions and Vehicles, Arcachon, France, July 2006

[9] D. Bonetti et al. Re-entry Mission Analysis of the Advanced Re-entry Vehicle (ARV). 7th Symposium on Aerothermodynamics for Space Vehicles, Brugges, Belgium, May 2011

[10] D. Bonetti et al. "Mission Analysis And GNC Of The Re-Entry Of The ARV Capsule". (IAC-12.D2.3.5) 63rd International Astronautical Congress, Naples, Italy, October 2012

[11] R. Haya Ramos, "Planetary Entry Toolbox: a SW suite for Planetary Entry Analysis and GNC Design" 2nd European Workshop on Astrodynamic Tools and Techniques, ESTEC, Noordwijk, September 2004

[12] R. Haya Ramos, and L. F. Peñín, "Planetary Entry Toolbox: a SW suite for Planetary Entry Design and Analysis" 3rd European Workshop on Astrodynamic Tools and Techniques, ESTEC, Noordwijk, October 2006

[13] M. Tauber, "A review of High-Speed, Convective, Heat-Transfer Computation Method", NASA Technical Paper 2914, 1989

[14] http://en.wikipedia.org/wiki/Chelyabinsk_event, February 2013

[15] http://ogleearth.com/2013/02/reconstructing-the-chelyabinsk-meteors-path-with-google-earth-youtube-and-high-school-math/, February 2013

[16] Borovicka, J., Spurny, P., and Shrbeny, L., Trajectory and orbit of the Chelyabinsk superbolide, Electronic Telegram No. 3423, Central Bureau for Astronomical Telegrams

[17] http://neo.jpl.nasa.gov/news/fireball_130301.html, 07 March 2013

[18] Zuluaga, J. I., and Ferrin, I. "A preliminary reconstruction of the orbit of the Chelyabinsk Meteoroid", arXiv:1302.5377v1, February 2013

[19] http://www.nasa.gov/mission_pages/asteroids/news/asteroid20130215.html, February 2013

[20] Zuluaga, J. I., Ferrin, I., and Geens, S. "The orbit of the Chelyabinsk event impactor as reconstructed from amateur and public footage", arXiv:1303.1796v1, March 2013

[21] http://ogleearth.com/2013/03/three-trajectory-models-of-the-chelyabinsk-meteoroid-compared/, March 2013

[22] http://english.ruvr.ru/photoalbum/106310514/106310537/

[23] https://maps.google.ee/maps/ms?msid=216221265233140305376.0004d5da6860954d651ba&msa=0&ll=55.013851,61.333923&spn=0.872465,2.458191, March 2013

[24] http://english.ruvr.ru/2013_02_18/Chelyabinsk-meteor-lake-to-become-a-must-see/